\documentclass[12pt,preprint]{aastex}

\begin{document}


\title{Modelling Solar Oscillation Power Spectra:  \\
    II. Parametric Model of {\bf Spectral Lines} Observed in Doppler Velocity Measurements}


\author{Sergei V. Vorontsov}
\affil{Astronomy Unit, School of Physics and Astronomy, Queen Mary University of London, Mile End
 Road, London E1 4NS, UK; S.V.Vorontsov@qmul.ac.uk \\
 Institute of Physics of the Earth, B. Gruzinskaya 10, Moscow
 123810, Russia}

\and

\author{Stuart M. Jefferies}
\affil{Institute for Astronomy, University of Hawaii, 34 Ohia Ku Street, Pukalani, HI 96768, USA;
stuartj@ifa.hawaii.edu \\
 Steward Observatory, University of Arizona, 933 North Cherry Avenue,
 Tucson, AZ 85721-0065, USA}





\begin{abstract}
We describe a global parametric model for the observed power spectra
of solar oscillations of intermediate and low degree.
A physically motivated parametrization is used as a
substitute for a direct description of mode excitation and damping
as these mechanisms remain poorly understood. The model
is targeted at the accurate fitting of power spectra coming from
Doppler velocity measurements and uses an adaptive response function that accounts for
both the vertical and horizontal components of the velocity field on
the solar surface and for possible instrumental and observational distortions.
The model is continuous in frequency, can easily be adapted to intensity measurements and extends naturally to the analysis of high-frequency pseudo modes (interference peaks at frequencies above the atmospheric acoustic cutoff).

\end{abstract}



\keywords{methods: data analysis --- Sun: helioseismology
--- Sun: oscillations}


\section{Introduction}
Helioseismic measurements of the deep solar interior are based on
the inversions of the solar oscillation frequencies and frequency splittings inferred from observations of the velocity or intensity field
on the visible part of the solar surface. With the exceptional quality and duration of the data now available from dedicated ground-based (GONG, BiSON) and space projects (SOHO MDI, GOLF, SDO HMI), it has become obvious
that progress in improving the accuracy and
resolution of helioseismic measurements is limited not by the
solar noise and finite mode lifetimes, but by the
current procedures for measuring the p-mode frequencies
and frequency splittings from the observed power spectra. The most damaging
aspect of current procedures is the
systematic errors they introduce; these are clearly seen both when comparing the
results obtained with different data analysis pipelines
\citep{schou02,basu03} and in the helioseismic inversions
where they manifest themselves as internal inconsistencies in the
input data sets \citep{svv02a,svv02b}.
Systematic errors can not be reduced by extending the duration of the observations and thus mask any gains in signal-to-noise ratio that come from the longer data set.

One source of problem which limits the accuracy of the frequency
measurements, is uncertainties in the spatial response function (also known as the leakage matrix) for the observations. The spatial response function
depends on the area of the Sun observed and on the details of the instrumentation. The latter can change with time and if the instrument is not accessible for comprehensive regular calibration \citep[one issue with SOHO MDI instrument is a gradual drift
of the focal distance, see e.g.][]{korz04}, can be particularly damaging. Fortunately, this problem can be mitigated
by using an adaptive response function \citep{vj05} (Paper I).

Another source of problems, which we address here,
is related to the modelling of the intrinsic line profiles.
Current (published) frequency measurements are based on fitting individual
lines with Lorentzian profiles. This is motivated by the analogy of solar oscillations with a simple damped
single harmonic oscillator. However, many of the spectral peaks in the solar oscillation power spectrum are asymmetric.
Although this asymmetry has typically been accommodated by adding an
extra asymmetry parameter (e. g. \citet{korz05}) to the Lorentzian model, there are downsides to this approach:
``asymmetric Lorenzian'' line profiles are only applicable in the
immediate vicinity of a sharp resonance and the addition of the extra parameter can reduce the stability of the frequency measurements.

Ideally, theoretical simulation of observational power spectra
would incorporate physical modelling of wave excitation and
damping. Unfortunately, this approach is not feasible due to our poor knowledge of the
turbulent convection and its interaction with waves in the outer
solar layers. This lack of knowledge, however, does not mean that
there isn't a way to provide a model suitable for helioseismic inversions that is based on some
physically-motivated parameterizations. There is an analogy here with a similar
difficulty which arises when the p-mode frequencies are used to
study the solar internal structure. In this case we have poor knowledge
of the structure of the near-surface layers and of the physics of the
wave behavior there. However, because the wave propagation in
these layers is very nearly vertical, it is reasonable to expect
that the near-surface effects change slowly with frequency and
depend on frequency only (at least when
the degree $l$ is not too high). This {\it a'priory}
knowledge allows us to separate the uncertain effects into a
frequency-dependent ``surface phase shift'' which as itself can
be used later as a separate diagnostic of the near-surface
layers. Without implementing this knowledge, which allows us to split
the inverse problem, we would not be able to infer anything about
the solar interior from the solar oscillation frequencies (more
discussion on this topic can be found in \citep{gv95}).

In developing our spectral model,
we will follow the very same logic and will split the problem in exactly the same manner.
We will make no attempt to consider the physics behind the mode
excitation and damping. Instead, we will only assume that the
two mechanisms operate close to the solar surface (an assumption
which is supported by the observations). For a
spherically-symmetric Sun, both the excitation strength and
damping shall change slowly with frequency (slow on a scale of
frequency separation between modes of consecutive radial order), and shall depend on frequency only when
the degree $l$ is not too high. At higher degree, just like in the
``surface phase shift'', the dependence on $l$ at constant frequency is expected to
emerge in proportion to $l(l+1)$, which governs the inclination of
the acoustic ray paths to vertical in the leading order.

This work on spectral modelling is a part of a wider project targeted at streamlining
the helioseismic inversion process (by analyzing the observational power
spectra directly, without frequency measurements) \citep{smj04}.
However, it can also serve the more limited goal of improving the accuracy of
frequency measurements. By allowing enough flexibility in the
spectral model we will substantially reduce systematic errors. In addition, by
reducing the total number of fitting parameters (when this is allowed by the data) we will improve the stability of the measurements and
reduce random errors.

Obviously, eliminating systematic errors without introducing new
ones is only possible if the model is physically relevant and the underlying basic assumptions are sound. There
are three tests that can be used to verify this. The first is how well the model fits the
observations. The second is whether the fitted parameters meet
the original expectations for their behavior (e.g. whether or not
the excitation amplitude falls on a single function of frequency
over a wide range in degree). The third and final test is helioseismic inversion
which is sensitive to the internal consistency of the input data
set (the measured frequencies).

 \section{The spectral model}
 We will develop our model in the temporal frequency domain.
 We assume the Sun to be spherically symmetric (deviations
 from spherical symmetry are usually treated as small perturbations), and expand all the variables in
our model in terms of spherical
 harmonics. We also assume the wave excitation comes from uncorrelated excitation
 events localized somewhere near the solar surface and are random in both time and in the
 angular coordinates over the solar surface: this makes their
 contributions to the observational signal uncorrelated, a
 property which comes from orthogonality of spherical harmonics.
 This in turn allows us to reduce the excitation problem to one of considering the response of
 the dynamical system to a single excitation event.

 We start by considering a simplified ideal measurement in which
 the entire solar surface is observed so that a particular
 spherical-harmonic component in the signal can be perfectly
 isolated. We also consider the signal, defined over
 the solar surface, as a scalar field (e.g. an intensity perturbation
 or radial velocity). Complications arising from the line-of-sight
 projection of the vector velocity field that occurs in realistic measurements, which are
 limited by the visible part of the solar surface, will be
 addressed at the end of this section.

 At this point it is convenient to first discuss the functional form of our spectral model, and the basic underlying assumptions, before we present a mathematical derivation.

 The general expression for the amplitude of a signal observed in
 a particular spherical harmonic is
 \begin{equation}
 {\cal A}_{\rm obs}=
 {{\cal A}_1e^{2i\theta}+{\cal A}_2\over1-Re^{2i\theta}}+{\cal B}_{\rm
 c}.
 \label{e1}
 \end{equation}
 All the parameters in the right-hand side of this expression are
 slowly-varying functions of frequency. $\theta$ is the phase
 integral of the trapped acoustic wave, a monotonically
 increasing function of frequency which takes values $\pi(n+1)$
 at frequencies of acoustic resonances ($n$ is the radial order).
 Energy dissipation is assumed to be localized in the sub-surface
 layers and the effects of energy losses are parameterized by the
 surface acoustic reflectivity of absolute value $R$. ${\cal A}_1$
 and ${\cal A}_2$ are two complex excitation amplitudes. The complex
 variable ${\cal B}_c$ designates the ``coherent component'' of
 the solar noise (or the ``direct visibility'' of the excitation
 source) which is thought to be responsible for the different
 sign of line asymmetries in intensity and velocity
 measurements \citep{rv97,nigam}. A random (uncorrelated)
 component of the solar noise will be added to the analysis later
 on (equation \ref{e19}).

 We provide two derivations of equation (\ref{e1}). The first
 derivation is based on the analysis of the linear adiabatic
 oscillation equations. We note that the excitation domain is
 not limited to the deeper layers, where the adiabatic approximation
 to the oscillation equations is thought to be valid, and can
 well extend to the near-surface layers where the physics of the
 oscillations is poorly understood. However, the general functional form of the equation (\ref{e1})
 does not change when the excitation operates partly from the
 non-adiabatic domain (as we will see from the second derivation which does not rely on the adiabatic approximation).

 Separation of the variables in the oscillation equations is performed in
 a standard way with the displacement field ${\bf u}$ and
 the Eulerian pressure perturbation $p'$ written as

 \begin{equation}
 {\bf u}({\bf r},t)=\left[
 U(r)\hat{\bf r}Y_{\ell m}(\theta,\phi)
 +V(r)\nabla_1Y_{\ell m}(\theta,\phi)\right]\,e^{-i\omega t}.
 \label{e2}
 \end{equation}
 \begin{equation}
 p'({\bf r},t)=p_1(r)Y_{\ell m}(\theta,\phi)\,e^{-i\omega t}.
 \label{e3}
 \end{equation}
 Here $\nabla_1=\hat{\bf\theta}(\partial/\partial\theta)
 +(1/\sin\theta)\hat{\bf\phi}(\partial/\partial\phi)$ is the
 angular part of the gradient operator and the hat symbol is used to designate
 unit vectors. The oscillation equations are implemented in
 the low-density envelope, where the Cowling approximation (which
 neglects perturbations of the gravitational field) is locally
 applicable. In this approximation the oscillation equations can be written as a single
 second-order equation
 \begin{equation}
 {d^2\psi\over d\tau^2}+
 \left[\omega^2-V(\tau)\right]\psi=0
 \label{e4}
 \end{equation}
 with
  \begin{equation}
 \psi=\rho_0^{-1/2}r
 \left({1\over c^2}-{\tilde w^2\over r^2}\right)^{1/4}
 \left(1-{N^2\over\omega^2}\right)^{-1/2}p_1
 \label{e5}
 \end{equation}
 and the independent variable $\tau$, defined as
  \begin{equation}
 {d\tau\over dr}={1\over c}
 \left(1-\tilde w^2{c^2\over r^2}\right)^{1/2},
 \label{e6}
 \end{equation}
where $\tilde w^2=l(l+1)/\omega^2$.
The acoustic potential $V$ is given by
 \begin{equation}
 V=N^2+\left[{r^2h\over\omega^2-N^2}\left({1\over c^2}-{\tilde w^2\over r^2}\right)^{1/2}\right]^{-1/2}
 {d^2\over d\tau^2}
 \left[{r^2h\over\omega^2-N^2}\left({1\over c^2}-{\tilde w^2\over r^2}\right)^{1/2}\right]^{1/2}
  \label{e7}
 \end{equation}
where $N$ is the Brunt-V\"ais\"ala frequency and
 \begin{equation}
 h=\exp\int\limits_0^r\left({N^2\over g_0}-{g_0\over c^2}\right)dr.
 \label{e8}
 \end{equation}
It is straightforward to show that the time-averaged energy flux, carried by an arbitrary local solution to the adiabatic oscillation equations in the upward direction, is
 \begin{equation}
 {i\over 4}\omega r^2\left(U^*p_1-Up_1^*\right)={i\over 4\omega}W\left(\psi,\psi^*\right),
 \label{e9}
 \end{equation}
where $W\left(\psi_1,\psi_2\right)$ is the Wronskian of the two solutions $\psi_1$  and $\psi_2$,
 \begin{equation}
 W\left(\psi_1,\psi_2\right)=\psi_1{d\psi_2\over d\tau}-\psi_2{d\psi_1\over d\tau},
 \label{e10}
 \end{equation}
and the star symbol designates complex conjugate.

We specify the external solution, which satisfies outer boundary conditions, by a sum of a wave incident on the surface and a reflected wave:
 \begin{equation}
 \psi_e={\cal A}_e\left(\psi_++R\psi_+^*\right),
 \label{e11}
 \end{equation}
 where $\psi_+$ is an upward wave, $\psi_+^*$ is a downward wave, $R$ is a real-valued reflectivity coefficient and ${\cal A}_e$ is a complex-valued function of frequency. The internal solution, which satisfies the inner boundary conditions, is specified as a standing wave
 \begin{equation}
 \psi_i={\cal A}_i\left(e^{i\theta}\psi_++e^{-i\theta}\psi_+^*\right)
 \label{e12}
 \end{equation}
 where $\theta$ is the phase integral across the acoustic cavity; when the reflectivity $R$ is set to 1, the two solutions match each other when $\theta=\pi(n+1)$ (eigensolutions of the adiabatic problem).

 To address the wave excitation, we add Dirac's $\delta$-function $\delta(\tau-\tau_s)$ to the right-hand side of equation (\ref{e4}). A particular solution of the resulting inhomogeneous equation is the Green's function
 \begin{equation}
 G(\tau,\tau_s)=\left\{
 -W^{-1}(\psi_+,\psi_+^*)\left[\psi_+(\tau_s)\psi_+^*(\tau)-\psi_+^*(\tau_s)\psi_+(\tau)\right], \quad \tau\le\tau_s
 \atop
 0, \quad \tau\ge\tau_s.
 \right.
 \label{e13}
 \end{equation}
 A general solution which satisfies the outer boundary conditions is $G(\tau,\tau_s)+\psi_e$, where $\psi_e$ is given by equation (\ref{e11}). By matching this solution with the internal solution $\psi_i$ (equation \ref{e12}), we obtain the amplitude of the external solution
 \begin{equation}
 {\cal A}_e=-W^{-1}(\psi_+,\psi_+^*)
 {\psi_+(\tau_s)e^{2i\theta}+\psi_+^*(\tau_s)\over 1-Re^{2i\theta}}.
 \label{e14}
 \end{equation}
We see that the right-hand side of the equation (\ref{e14}) fits the functional form of the right-hand side of equation \ref{e1}. Since an arbitrary excitation source can be considered as a linear superposition of ``elementary'' excitations described by $\delta$-functions, it will bring an observational signal of the functional form specified by equation (\ref{e1}).

An alternative derivation comes from considering wave propagation and interference in the acoustic cavity in analogy with light propagation in a Fabry-Perot interferometer \citep[e.g.,][]{duvall93}. Consider two waves emitted by an excitation source which is localized near the surface: an upward wave and a downward wave. The upward wave is reflected almost immediately from the surface and then undergoes multiple further reflections from the surface as it travels up and down across the acoustic cavity. The signal observed at the surface is evaluated as a sum of a geometric series with the common ratio $Re^{2i\theta}$, where $2\theta$ is the phase lag between two consecutive reflections from the surface. This signal gives the term ${\cal A}_2/(1-Re^{2i\theta})$ in equation (\ref{e1}). The other term in equation (\ref{e1}), ${\cal A}_1e^{2i\theta}/(1-Re^{2i\theta})$, comes from the wave which is emitted downwards and which accumulates an additional phase lag of $2\theta$ before reaching the surface.

To obtain a convenient expression for the signal power $|{\cal A}_{\rm obs}|^2$, we introduce the variable transformation
\begin{equation}
 \tan\theta={1-R\over 1+R}\tan\varphi,
 \label{e15}
 \end{equation}
which gives
\begin{equation}
 {1\over 1-Re^{2i\theta}}={1+Re^{2i\varphi}\over 1-R^2},\quad {e^{2i\theta}\over 1-Re^{2i\theta}}={R+e^{2i\varphi}\over 1-R^2}.
 \label{e16}
 \end{equation}
 Equation (\ref{e1}) can thus be written as
\begin{equation}
 {\cal A}_{\rm obs}={b+ae^{2i\varphi}\over 1-R^2}
 \label{e17}
 \end{equation}
 with
\begin{equation}
 b=R{\cal A}_1+{\cal A}_2+\left(1-R^2\right){\cal B}_c,\quad a={\cal A}_1+R{\cal A}_2,
 \label{e18}
 \end{equation}
 and we have
  \begin{equation}
 \left|{\cal A}_{\rm obs}\right|^2
 =\left[{A\cos\left(\varphi-S\right)\over 1-R^2}\right]^2
 +B^2,
 \label{e19}
 \end{equation}
where
 \begin{equation}
 A^2=4\left|ab\right|,
 \label{e20}
 \end{equation}
 \begin{equation}
 B^2=\left(|a|-|b|\over 1-R^2\right)^2+B_u^2,
 \label{e21}
 \end{equation}
 \begin{equation}
 2S=\arg b -\arg a,
 \label{e22}
 \end{equation}
 and we have added an uncorrelated component of the solar background $B_u^2$, to the right-hand side of equation (\ref{e19}).

 We now designate $L_U$ and $L_V$ as the two components of the response function (leakage matrix) which specify the sensitivity of the spatial filter of the Doppler-velocity measurements to the vertical and horizontal velocity components, respectively.  The observational power spectrum is then given by
 \begin{equation}
 P=\left|L_U+hL_V\right|^2
 \left\{\left[{A\cos\left(\varphi-S\right)\over 1-R^2}\right]^2
 +B^2\right\}
 \label{e23},
 \end{equation}
 where $h$ is the ratio of horizontal and vertical components of the displacement field on the solar surface ($V/U$, see equation \ref{e2}). A simple theoretical prediction for this quantity is
  \begin{equation}
 h={GM\over R^3\omega^2}.
 \label{e24}
 \end{equation}
 This value comes from the horizontal component of the momentum equation in the Cowling approximation, when the Lagrangian pressure perturbation is zero, see e.g. \citep[][]{cox80}; it is thus applicable to any divergence-free flow at any frequency, not only at the frequencies of acoustic resonances. Our practical experience with fitting solar velocity power spectra provides no evidence that this ratio shall be modified, a conclusion that is in agreement with \citet{rhodes01}. We note, however, that a small variation of $h$ can not be distinguished from a small distortion of the leakage matrix caused by an error in the plate scale (Paper I).
 Discussions of other efforts of measuring $h$ can be found in \citep{rhodes01} and \citep{korz04}.

 Equation (\ref{e23}) represents the spectral model which we use for matching the observed power spectra using an iterative maximum-likelihood technique. At each degree $l$, the parameters $A,R,S,B$ are expected to change slowly with frequency (we discard their possible variation with azimuthal order $m$ in the results presented below). $A$ is an effective excitation amplitude and $R$ is the acoustic reflectivity of the solar surface. These two parameters replace the mode amplitude and line width which are used in traditional spectral models for the oscillation modes. Since the excitation and damping mechanisms are thought to operate in the near-surface layers, where wave propagation is nearly vertical, we expect $A$ and $R$ to depend on frequency only (not on the degree), at least when the degree $l$ is not too high. The parameter $S$ specifies line asymmetry: its relation with ``asymmetric Lorentzian'' line profiles is discussed in Appendix A. $B$ is a composite background with contributions from both correlated and uncorrelated components of the solar noise. The functional form of equation (\ref{e23}) assumes that horizontal and vertical components of the background are in the same ratio as those of the resonant signal. This is hardly the case for solar granulation noise. The inaccuracy in this assumption leads to a distortion of the measured values of $B^2$. We will get back to this point later when discussing our numerical results.

 For a spherically-symmetric configuration, the positions of the spectral resonances (as well as the line profiles at given  $R$ and $S$), are governed by the phase integral $\theta(\omega)$ which has to be specified in the spectral model for each value of degree $l$. Ideally, $\theta(\omega)$ is computed from an equilibrium solar model which fits the seismic data. As such a model is not available (it is itself a result of helioseismic structural inversion of the measured oscillation frequencies), we use an iterative approach where $\theta(\omega)$ is parameterized by the positions of the spectral resonances $\omega_n$, where $\theta(\omega_n)=\pi(n+1)$. A continuous variation of $\theta(\omega)$  between the resonances is represented by a spline with prescribed gradients $d\theta/d\omega$ at the resonant frequencies $\omega=\omega_n$. The values of these gradients are evaluated using solutions to the adiabatic oscillation equations computed for a proxy solar model. Using the unrestricted variational principle described in \citet{svv13}, it is straightforward to show that
  \begin{equation}
 {1\over\omega^2}W\left(\psi_i,\psi_e\right)=\delta\left(\omega^2\right)
 \int\limits_0^R\rho_or^2\left[U^2+l(l+1)V^2\right]dr.
 \label{e25}
 \end{equation}
 In this equation $\psi_i$ is the solution to the adiabatic oscillation equations which satisfies the inner boundary conditions, $\psi_e$ is the solution which satisfies the (conservative) outer boundary conditions and $\delta(\omega^2)=\omega^2-\omega_n^2$. Using this variational principle, we have
  \begin{equation}
  {d\theta\over d\omega}\Big|_{\omega=\omega_n}=
  {2\omega^2\over\psi^2+{1\over\omega^2}\left({d\psi\over d\tau}\right)^2}
  \int\limits_0^R\rho_or^2\left[U^2+l(l+1)V^2\right]dr,
 \label{e26}
 \end{equation}
 with $\psi$ and $d\psi/d\tau$ evaluated at the depth where $V(\tau)\ll\omega^2$ (see equation \ref{e4}).
 We note that due to the high-frequency asymptotic properties of solar p modes, the functions $\theta(\omega)$ are close to linear functions of frequency $\omega$. A small inaccuracy in the gradients $d\theta/d\omega$ leads only to a small distortion of the measured values of the acoustic reflectivity $R$ (see equation \ref{e15}: the line width in the power spectrum depends on both $R$ and  $d\theta/d\omega$).

 For the simple scenario where $A, R$ and $B$ do not depend on frequency, $\theta$ changes linearly with $\omega$, and $S=0$, the power spectrum predicted by equation (19) can be represented by a sum of Lorentzian profiles (see Appendix A).

Using equations (A8, A9) we find that the line width at half power, measured in units of angular frequency $\omega$, is
 \begin{equation}
 \Lambda=-2\gamma=-{\ln R\over d\theta /d\omega}.
 \label{e27}
 \end{equation}
 This relation shows that even when $R$ is degree-independent, the line width depends explicitly on the degree $l$, being inversely proportional to the ``mode mass'' (the integral in equation \ref{e26} is proportional to mode energy).

 The oscillation frequencies and frequency-splitting coefficients are measured using an iterative procedure.
 Our preliminary analysis (with results presented in the next section) was targeted at measuring the parameters $A,R,S,B$ for individual modes
 to check if they reveal the expected behavior (a variation with frequency only, at least when the degree $l$ is not too high). For this reason,
 the adjustments of the frequencies and splittings were alternated with improvements of $A,R,S$ and $B$ as functions of frequency at given $l$. A dependence of $A,R,S,B$ on frequency inside the fitting domains
  was allowed for by adding a uniform shift to a single (piecewise linear) function of frequency. This function describes the average variation of the corresponding parameter with frequency for lower-degree modes (updated in the outer iteration after the inner iterations converge). In the results presented below, the size of the fitting domain in frequency was taken to be 10 linewidths
for measuring $A$ and $R$, an average of the frequency spacings between modes of consecutive order when measuring $S$ and $B$, and about 6 $\mu$Hz when measuring the frequencies and frequency splittings.
 Separate tests where performed to verify that the fitted parameters do not suffer from systematic variations when the sizes of the fitting intervals change.

 The leakage matrix was computed as described in Paper I with two extensions. One is a full description of the effects due to a non-zero solar B angle (in Paper I these effects were only allowed as a perturbation).  This extension to the leakage-matrix computation is described in Appendix B. The other extension is a more accurate treatment of the mode coupling by differential rotation, which is now performed as described in \citep{svv07}. Spatial leaks in the range of $\pm 30$ in $l$ and $m$ where included in the spectral modeling. The large number of leaks had to be taken into account for addressing $B$ at frequencies higher than about 2 mHz, where the solar background appears to be buried below the resonant signals of spatial leaks.

 We note that the spectral model described by equation (23) is applicable not only to acoustic resonances, but also to the high-frequency interference pikes, or pseudo modes ($R\rightarrow 0$) observed at frequencies higher than the ``acoustic cutoff'' frequency (about 5 mHz). In addition, adaptation of the spectral model to observations in intensity just requires a modification of the response function ($L_U+hL_V$ in equation 23).

 \section{Numerical results}
 A representative example of the 1-year SOHO MDI power spectrum, obtained from the first year of observations (at low solar activity), and its model is shown in Fig. 1. Here the spectra corresponding to different values of azimuthal order $m$ have been shifted in frequency and then averaged for better visual comparison. The observed complexity of the asymmetric spectral lines is due to the contribution of multiple spectral leaks. We note the noise level is difficult to detect as it is below the amplitudes of resonant signals.

  The parameters of the spectral model are illustrated in Figs 2--5. The results are shown for $p$ modes of radial order $n$ from 1 to 10 and for the $f$ modes of degree $l\leq 200$. The maximum-likelihood solution was obtained by an iterative improvement of both the spectral parameters $(A, R, S, B)$ and the resonant frequencies and frequency splittings. The resulting agreement between the model and the data is satisfactory, as can be judged from a proper merit function \citep{anderson90} and from the visual inspection of the $m$-averaged power spectra. However, a small systematic inaccuracy in the predicted amplitudes of the spatial leaks remains. The origin of this mismatch is not yet understood; it may be related, in part, to the asymmetric distortion of the point-spread function of the MDI instrument.

The spectral parameters of individual modes do not depend on the degree and collapse to slowly-varying functions of frequency only, when the degree $l$ is not too high (less than about 100). The composite background $B^2$ (Fig. 5) may look to be an exception to this. However, an accurate measurement of the background at frequencies higher than about 2 mHz is a difficult task as the background level appears to be significantly smaller than the resonant signals coming from the spatial leaks. As a result, the measurement of $B^2$ can be distorted by small inaccuracies in the leakage matrix. The fitted background is also significantly higher than the average at the lowest values of degree $l$ (e.g. $p_{10}$ mode of $l=2$ in Fig. 5). A likely explanation of this excess is instrument-related noise due to variations in the exposure times for the series of observations acquired at different wavelengths by MDI and used to compute the solar velocity signal \citep{schou13}. This noise can probably be modeled by adding an $l=0$ component to the (otherwise degree-independent) solar background $B^2(\omega)$.

Interestingly, the analysis of solar $f$ modes reveal the same values of the excitation amplitudes $A$ and ``acoustic reflectivity'' $R$ as solar $p$ modes of similar frequencies (Figs 2, 3). This indicates that the excitation and damping mechanisms do not distinguish between $p$- and $f$ modes, despite the difference in their physical nature (the $f$ modes are incompressible waves). The composite background of $f$ modes appears to be smaller than that of $p$ modes (Fig. 5). Since the fitted background $B^2$ at low frequencies is dominated by the granulation noise, one possible explanation is that its contribution to the observational power shall be modeled with a smaller (or zero) value of $h$, which specifies the ratio of horizontal and vertical velocities. When processing the data we used a simple theoretical value (equation \ref{e24}) which corresponds to incompressible adiabatic motion. This approximation is hardly relevant to granulation noise.

Fig. 6 shows the rotational splitting coefficients resulting from the measurement process described above, in comparison with published splitting coefficients \citep{schou99}. The horn-like structures in the published results, which signify systematic errors, are eliminated. As indicated by a detailed analysis \citep{svv09}, the dominant part of the systematic errors came from discarding the effects of mode coupling by differential rotation in the original version of the SOHO MDI data analysis pipeline \citep[which was later improved to include these effects, among others; see][ who also observed a reduction  in the horn-like structures]{larson08}.

The differences between the centroid frequencies and their published values are shown in Fig. 7. The major part of the discrepancies is due to the line asymmetries which were not accounted for in the original version of the MDI data analysis pipeline. Smaller-scale features are due apparently to the combination of the mode-coupling effects with the plate-scale error. The centroid frequencies measured in this work have been used in a recent study targeted at the seismic diagnostics of the equation of state \citep{svv13}. Interestingly, it was found that these frequencies allow us to achieve a significantly better agreement with solar models. This is quite an unusual finding: better accuracy of observational data brings better agreement with theoretical models.

 \section{Discussion}
 When tested with high-quality 1-year SOHO MDI Doppler velocity measurements, the spectral model suggested in this paper allows an accurate description of the observational power spectra over a wide (continuous) range of frequency and in the degree range up to $l$ of about 200. The oscillation frequencies and frequency splittings measured by the new technique appear to have smaller systematic errors when compared with published results.

 A distinctive feature of the technique is its potential ability to reduce the random errors in the frequency- and frequency-splitting measurements of the p modes which penetrate into the deep solar interior. This benefit comes from the global approximations of the spectral parameters of the model (excitation amplitude $A$, acoustic reflectivity $R$, line asymmetry $S$ and composite background $B$). At both low and intermediate degree $l$ these parameters can be represented by slowly-varying functions of frequency only (a possible exception is the solar background $B$, which requires further study). When inferred from the large volume of high-quality intermediate-degree data, these global approximations can be used in measurements at lower degree $l$. The subsequent reduction in the number of free parameters in the mode-fitting algorithm is expected to bring significant improvement to the accuracy and precision of the frequency- and frequency-splitting measurements in this important region of the solar oscillation power spectrum (which contains information on the solar core).

 A problem which is not yet resolved is that the performance of the technique degrades when analyzing the MDI data of degree higher than about 200. This is apparently due to an inaccuracy in the predicted amplitudes of the spatial leaks. This leads to poor stability (and unavoidable systematic errors) of the frequencies and frequency splittings at higher degree where the spatial leaks start to blend into a single ridge. The origin of this problem may be related, at least in part, with the asymmetric distortion of the point-spread function of the MDI instrument \citep[see e.g.][]{korz04,korz13} which is not accounted for in our model.

 Our future work is targeted at processing the entire volume of the SOHO MDI data accumulated over the 15 years of the MDI operational lifetime, as well as implementation of our technique to the analysis of SDO HMI data. On the theoretical side, the ultimate goal of the project is a streamlined helioseismic inversion. Here frequency measurements will be eliminated from the analysis, and the parameters of the rotating and aspherical solar model will be matched directly with p-mode power spectra. This approach will bring the benefits of streamlined regularization by eliminating the problems associated with error correlation and possible mode misidentification in the frequency measurements.




\acknowledgments

We thank Jesper Schou and Tim Larson for providing Doppler-velocity power spectra from their analysis of SOHO MDI data, and for illuminating discussions of possible instrumental and observational distortions in these measurements. We also thank an anonymous referee for suggestions which helped to improve the presentation. SOHO is a project of international cooperation between ESA and NASA. This work was supported in part by the UK STFC under grant PP/E001459/1.



\appendix

\section{Eigenfunction expansion and line asymmetry}
 Using the identity
 \begin{equation}
 e^{\pm i\theta}=R^{\mp 1/2}\left[\cos\left(\theta-{i\over 2}\ln R\right)\pm i\sin\left(\theta-{i\over 2}\ln R\right)\right],
 \label{a1}
\end{equation}
equation (\ref{e1}), which describes the amplitude spectra, can be written as
 \begin{equation}
 {\cal A}_{\rm obs}={i\over 2}\left({1\over R}{\cal A}_1+{\cal A}_2\right)\cot\left(\theta-{i\over 2}\ln R\right)
 -{1\over 2}\left({1\over R}{\cal A}_1-{\cal A}_2\right)+{\cal B}_c.
 \label{a2}
 \end{equation}
 Introducing the independent variable
 \begin{equation}
 z=\theta-{i\over 2}\ln R,
 \label{a3}
 \end{equation}
 which is now a complex quantity, and expanding $\cot z$ in partial fractions, we have
  \begin{equation}
 {\cal A}_{\rm obs}={1\over 2}\left({1\over R}{\cal A}_1+{\cal A}_2\right)
 \left[{i\over z}+\sum\limits_{k=1}^\infty\left({i\over z-\pi k}+{i\over z+\pi k}\right)\right]
 -{1\over 2}\left({1\over R}{\cal A}_1-{\cal A}_2\right)+{\cal B}_c.
 \label{a4}
 \end{equation}
The right-hand side of this equation has simple poles at $z=\pm\pi k,\,\,\,k=n+1$, and describes ${\cal A}_{\rm obs}$ as being produced by a superposition of damped harmonic oscillations of consecutive radial orders. Since the complex amplitudes ${\cal A}_1$ and ${\cal A}_2$ are common for all the modes, the oscillations are excited in a coherent way and the resultant power spectrum can not be represented, in general, by a simple sum of power spectra of individual modes (modes described by the same spherical harmonic can not be considered as uncorrelated).
It can also be seen from equation (A4) that when ${\cal A}_1, {\cal A}_2$ and $R$ are constants, $\theta$ changes linearly with $\omega$, ${\cal B}=0$ and the contribution of neighboring resonances is discarded, an individual resonant line in the power spectrum $|{\cal A}_{\rm obs}|^2$ has a symmetric Lorentzian profile only if $\arg({\cal A}_1/R+{\cal A}_2)=\arg({\cal A}_1/R-{\cal A}_2)$, which is equivalent to $\arg({\cal A}_1)=\arg({\cal A}_2)$.

To address the profile of the resultant power spectrum, we perform a similar decomposition but for the observational power $|{\cal A}_{\rm obs}|^2$ (equation \ref{e19}). Using equation (\ref{e15}), we have
 \begin{equation}
 |{\cal A}_{\rm obs}|^2=A^2\left[{\cos 2S\over 8R^2}\,
 {1+{2R\over 1-R^2}\tan 2S\sin 2\theta\over {1+R^2\over 2R}-\cos 2\theta}
 -{1\over (1-R^2)^2}\left({1+R^2\over 4R}\cos 2S-{1\over 2}\right)\right]+B^2.
 \label{a5}
 \end{equation}
 Using $(1+R^2)/(2R)=\cosh\ln R,\,\,\,(1-R^2)/(2R)=\sinh\ln R$, and the standard expression for wrapped Cauchy distribution
 \begin{equation}
 \sum\limits_{k=-\infty}^\infty
 {\beta\over\pi\left[\beta^2+(\alpha-\mu+2\pi k)^2\right]}=
 {1\over 2\pi}{\sinh\beta\over\cosh\beta-\cos(\alpha-\mu)},
 \label{a6}
 \end{equation}
 we arrive to the desired expansion
  \begin{eqnarray}
 \nonumber |{\cal A}_{\rm obs}|^2&=&A^2\Big[-{\ln R\cos 2S\over 8R(1-R^2)}
 \left(1+{2R\over 1-R^2}\tan 2S\sin 2\theta\right)
 \sum\limits_{k=-\infty}^\infty{1\over(\theta-\pi k)^2+{1\over 4}\ln^2R}\\
 &-&{1\over(1-R^2)^2}\left({1+R^2\over 4R}\cos 2S-{1\over 2}\right)\Big]+B^2.
 \label{a7}
 \end{eqnarray}
 Consider a simple scenario when $A,R,S,B$ do not depend on frequency $\omega$, and $\theta$ is proportional to $\omega$, with $\theta=\pi\omega/\Delta\omega$, so that the resonant frequencies are equidistant. Discarding normalization and additive constants, the power spectrum is proportional to
 \begin{equation}
 \left[1+{2R\over 1-R^2}\tan 2S\sin\left({2\pi\omega\over\Delta\omega}\right)\right]
 \sum\limits_{k=-\infty}^\infty{1\over(\omega-k\Delta\omega)^2+\gamma^2}
 \label{a8}
 \end{equation}
 with
 \begin{equation}
 \gamma^2=\left({\Delta\omega\over 2\pi}\ln R\right)^2.
 \label{a9}
 \end{equation}
 When $S=0$, the power spectrum is a simple sum of Lorentzian profiles, an unexpected conclusion. When the asymmetry parameter differs from zero, the spectrum is modulated with a harmonic function of frequency with period equal to the frequency spacing $\Delta\omega$.

 In current work in solar seismology, a popular functional form for the asymmetric line profiles of solar p modes is that suggested by \citet{nigamkosovichev}:
 \begin{equation}
 {(1+Bx)^2+B^2\over 1+x^2},
 \label{a10}
 \end{equation}
 where $x=(\omega-\omega_n)/\gamma$; this expression was obtained by using a power-series expansion at small $\omega-\omega_n$ and hence is only applicable in the vicinity of the resonant frequency. One obvious problem with this profile it that when $x\rightarrow\pm\infty$, the power density tends to a non-zero value of $B^2$, which corresponds to an infinite mode energy. Matching the line profiles predicted by expressions (\ref{a8}) and (\ref{a10}) in the vicinity of the resonance, we have $S\simeq B$. \citet{korz04} suggested modification to the Nigam \& Kosovichev's profile of the form
 \begin{equation}
 {1+\alpha\left(x-\alpha/2\right)\over 1+x^2}.
 \label{a11}
 \end{equation}
 This profile brings negative power density when $x\rightarrow\infty$ or $x\rightarrow -\infty$, depending on the sign of the asymmetry parameter $\alpha$.
 Matching our result with this profile in the vicinity of $x=0$, we have $S\simeq\alpha/2$.

 Another approach to modeling observational signal over a continuous frequency range (all the way between the consecutive acoustic resonances) was suggested recently by \citet{hindman11}. The expression suggested for the observational amplitude differs markedly from our equation (A4): instead of a single frequency-dependent function ${\cal A}_1/R+{\cal A}_2$ which defines the normalization of all the partial fractions $i/(z-\pi k)$ (equation A4), each partial fraction enters the summation with individual complex amplitude. Line asymmetry in the power spectrum is interpreted in \citep{hindman11} as a result of mode interference (cross-terms between partial fractions describing individual damped harmonic oscillations). In our view, the origin of the line asymmetry is very different: when ${\cal B}_c=0$, the asymmetry comes from the emission, by the same excitation event, of both an upward- and downward propagating waves; the sign of the asymmetry and its strength are thus governed by the parity and depth of the excitation source \citep[see e.g.][and references therein]{rv95,rv97}. When observations are interpreted in terms of damped harmonic oscillations, this vision can be supported by a simple illustrative example. Consider an instantaneous excitation of a single mode of frequency $\omega_0$ and damping rate $\gamma$, followed by another similar excitation  after a short time interval $\Delta t$ specified by the temporal behavior of the excitation event. The combined power spectrum is proportional to
  \begin{equation}
  \left|{1\over\omega-\omega_0-i\gamma}+{e^{-i\omega\Delta t}\over\omega-\omega_0-i\gamma}\right|^2=
  {2\left(1+\cos\omega\Delta t\right)\over (\omega-\omega_0)^2+\gamma^2};
 \label{a12}
 \end{equation}
 and the spectral line is asymmetric when $\omega_0\Delta t\ne\pi k$, where $k$ is an integer, i.e. when the two excitations are not exactly in phase (or in counter-phase). The pair of consecutive excitations can be interpreted as being produced by the upward and the downward wave. It is seen from this example that line asymmetry is governed directly by the spatio-temporal properties of a single excitation event and does not require interference with modes of neighboring frequencies. Interpretation of line asymmetry it terms of mode interference also faces difficulty when the frequency separation $\Delta\omega$ is much bigger than the linewidth, i.e. when the amplitude of a neighboring mode is negligibly small in the vicinity of a sharp resonance.

 \section{Static part of the leakage matrix: generalization to $\beta\ne\pi/2$ rotation}
This appendix shall be read in conjunction with Paper I; it contains the replacement to section 2.2 of Paper I, to allow an explicit account of the leakage-matrix computation for non-zero solar B angle. The rotation of the coordinate system is now performed with $\beta=\pi/2+B$. Equations (25,26) of Paper I are replaced with more general transformation relations
 \begin{equation}
 Y_{lm}(\theta',\phi')=\sum\limits_{m'}d_{mm'}^{(l)}(\beta)Y_{lm'}(\theta,\phi),
 \label{b1}
 \end{equation}
 \begin{equation}
 Y_{lm}(\theta,\phi)=\sum\limits_{m'}d_{m'm}^{(l)}(\beta)Y_{lm'}(\theta',\phi'),
 \label{b2}
 \end{equation}
 with transformation coefficients
 \begin{equation}
 d_{m'm}^{(l)}(\beta)=\sum\limits_t{(-1)^t\sqrt{(l+m)!(l-m)!(l+m')!(l-m')!}\over(l-m'-t)!(l+m-t)!t!(t+m'-m)!}
 \left(\cos{\beta\over 2}\right)^{2l+m-m'-2t}\left(\sin{\beta\over 2}\right)^{2t+m'-m}
 \label{b3}
 \end{equation}
 where $t$ is run over those integer values for which all the factorials are those of positive integers or zero (cf equations A1, A2 of Paper I);  the transformation coefficients satisfy symmetry relations
 \begin{equation}
 d_{m'm}^{(l)}(\beta)=(-1)^{m+m'}d_{mm'}^{(l)}(\beta)=d_{-m,-m'}^{(l)}(\beta),
 \label{b4}
 \end{equation}
 \begin{equation}
 d_{m'm}^{(l)}(\pi-\beta)=(-1)^{l+m'}d_{m',-m}^{(l)}(\beta).
 \label{b5}
 \end{equation}
 The recurrence relations, convenient for calculating the transformation coefficients, are available in \citep{varshalovich}. After proper sign corrections \citep[due a transposed definition of the expansion coefficients, cf equation 5.5.1 of][]{varshalovich}, these relations are \citep[equations 4.8.16, 4.8.17 of][]{varshalovich}
 \begin{eqnarray}
 \sin\beta\left[(l+m)(l-m+1)\right]^{1/2}d_{m',m-1}^{(l)}(\beta)&-&
 2(m'-m\cos\beta)d_{m'm}^{(l)}(\beta)\\
 \nonumber &+&\sin\beta\left[(l+m+1)(l-m)\right]^{1/2}d_{m',m+1}^{(l)}(\beta)=0,
 \label{b6}
 \end{eqnarray}
 \begin{eqnarray}
 \sin\beta\left[(l+m)(l-m+1)\right]^{1/2}d_{m-1,m'}^{(l)}(\beta)&+&
 2(m'-m\cos\beta)d_{m,m'}^{(l)}(\beta)\\
 \nonumber &+&\sin\beta\left[(l+m+1)(l-m)\right]^{1/2}d_{m+1,m'}^{(l)}(\beta)=0.
 \label{b7}
 \end{eqnarray}
 Due to symmetry relations (\ref{b4}), only a quarter of the matrix of transformation coefficients at each $l$ needs to be evaluated using the recurrence relations; when equation (B7) is in use the recurrence starts at $m'=l$ using the explicit expressions
  \begin{equation}
 d_{lm}^{(l)}(\beta)=\left[{(2l)!\over(l+m)!(l-m)!}\right]^{1/2}
 \left(\cos{\beta\over 2}\right)^{l+m}\left(\sin{\beta\over 2}\right)^{l-m}.
 \label{b8}
 \end{equation}
 The static part of the leakage matrix is
 \begin{equation}
 S_{l'l}^{m'm}(\beta)=
 \sum\limits_{\mu=-l}^l\sum\limits_{\mu'=-l'}^{l'}
 d_{\mu m}^{(l)}(\beta)d_{\mu' m'}^{(l')}(\beta)C_{l'l}^{\mu'\mu}.
 \label{b9}
 \end{equation}
This expression replaces equation (35) of Paper I. Deviation of $\beta$ from $\pi/2$ (non-zero B angle) brings ``prohibited'' leaks (with $l+l'+m+m'$ odd). The static part of the leakage matrix obeys symmetry properties
 \begin{equation}
 S_{l'l}^{-m',-m}(\beta)=(-1)^{m+m'}S_{l'l}^{m'm}(\beta),
 \label{b10}
 \end{equation}
 \begin{equation}
 S_{l'l}^{m'm}(\pi-\beta)=(-1)^{l+l'+m+m'}S_{l'l}^{m'm}(\beta).
 \label{b11}
 \end{equation}
 It can be seen from the last relation that the ``prohibited'' leaks are zero when $\beta=\pi/2$. At small $B=\beta-\pi/2$, the amplitudes of ``prohibited'' leaks are proportional to $B$, and variations of the amplitudes of ``unprohibited'' leaks is quadratic in $B$. When analyzing the 1-year power spectra, we thus calculate the leakage matrix with $B=5.11$ degrees which gives a proper annual average for $B^2$.

 Using orthogonality and normalization properties of spherical harmonics, it can be seen that
 \begin{equation}
 \sum\limits_{m=-l}^ld_{pm}^{(l)}(\beta)d_{qm}^{(l)}(\beta)=\delta_{pq};
 \label{b12}
 \end{equation}
 this relation gives
  \begin{equation}
 \sum\limits_{m=-l}^l\sum\limits_{m'=-l'}^{l'}\left(S_{l'l}^{m'm}(\beta)\right)^2=
 \sum\limits_{m=-l}^l\sum\limits_{m'=-l'}^{l'}\left(C_{l'l}^{m'm}\right)^2,
 \label{b13}
 \end{equation}
 which means that the total power of leaks with given $l'$ to target $l$ does not depend on the inclination angle $\beta$.




\clearpage



 \begin{figure}
 \epsscale{1}
 \plotone{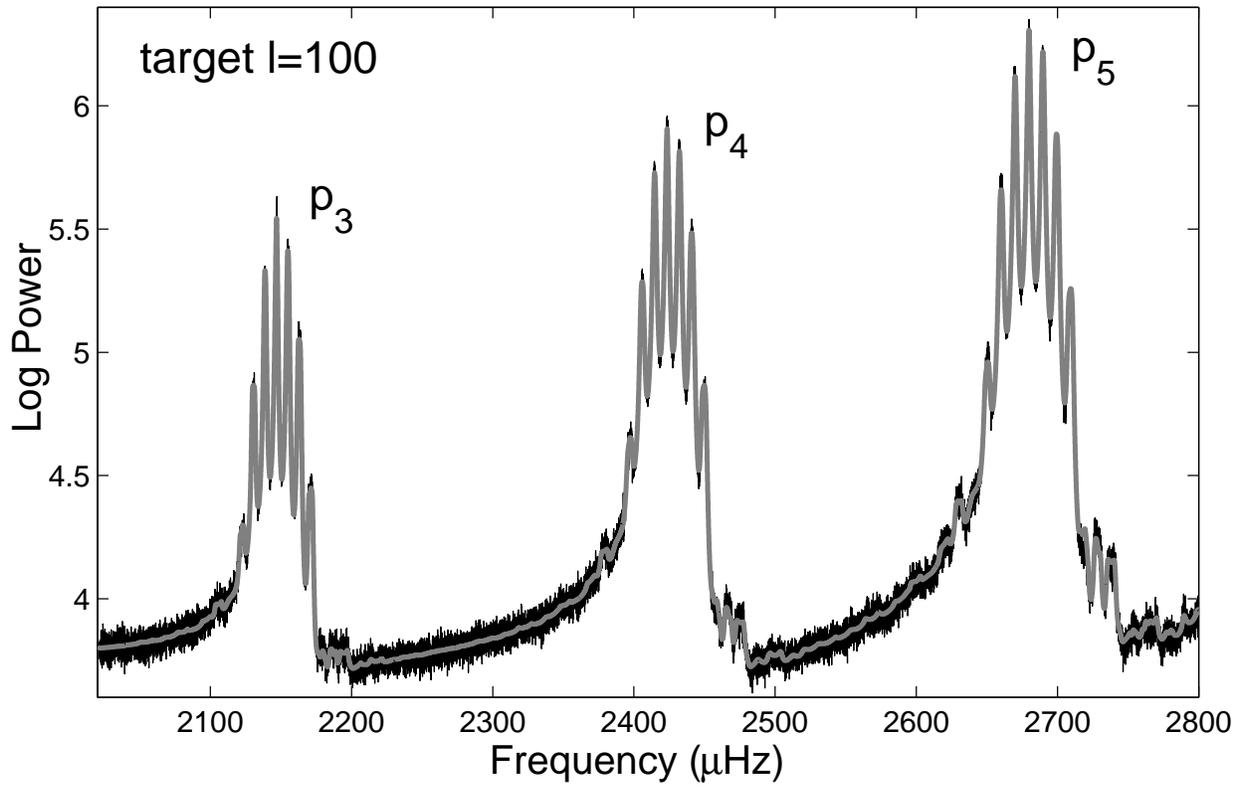} \caption{
 Part of the 1-year SOHO MDI $m$-averaged power spectrum at $l=100$ (thin line) and its model (thick gray line).
 \label{fig1}}
 \end{figure}
 \begin{figure}
 \epsscale{1}
 \plotone{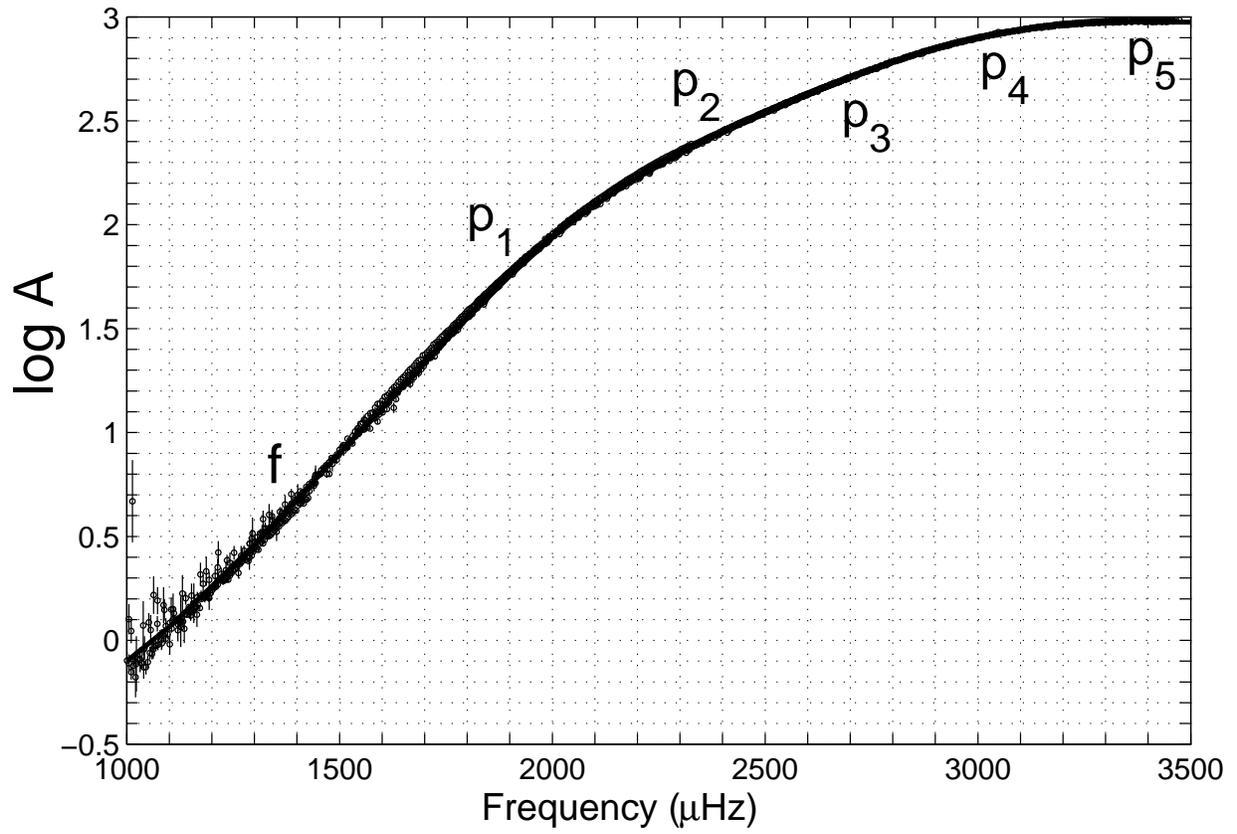} \caption{
 Excitation amplitude obtained from the 1-year SOHO MDI Doppler-velocity power spectra. Solid lines show approximation by a slowly-varying function of frequency for lower-degree modes.
 \label{fig2}}
 \end{figure}
 \begin{figure}
 \epsscale{1}
 \plotone{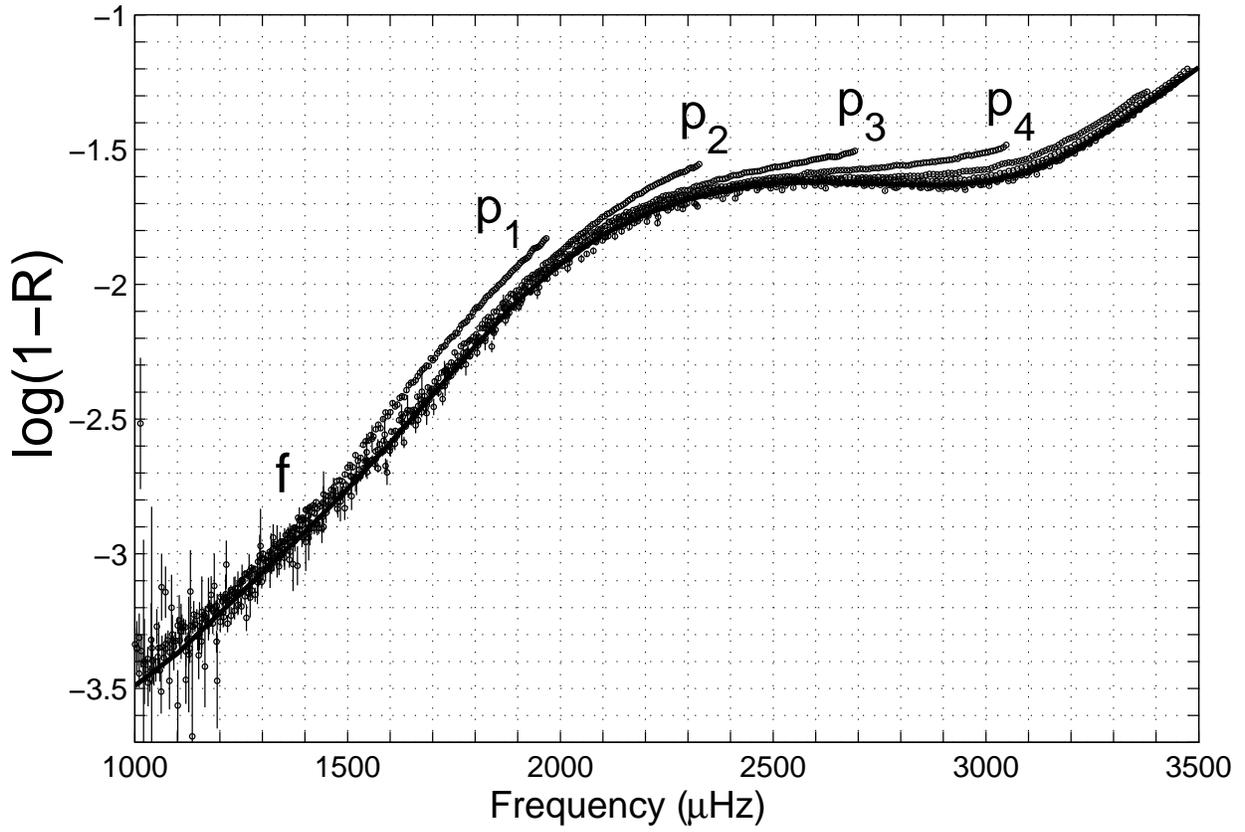} \caption{
 Same as Fig. 2, but for acoustic reflectivity.
 \label{fig3}}
 \end{figure}
 \begin{figure}
 \epsscale{1}
 \plotone{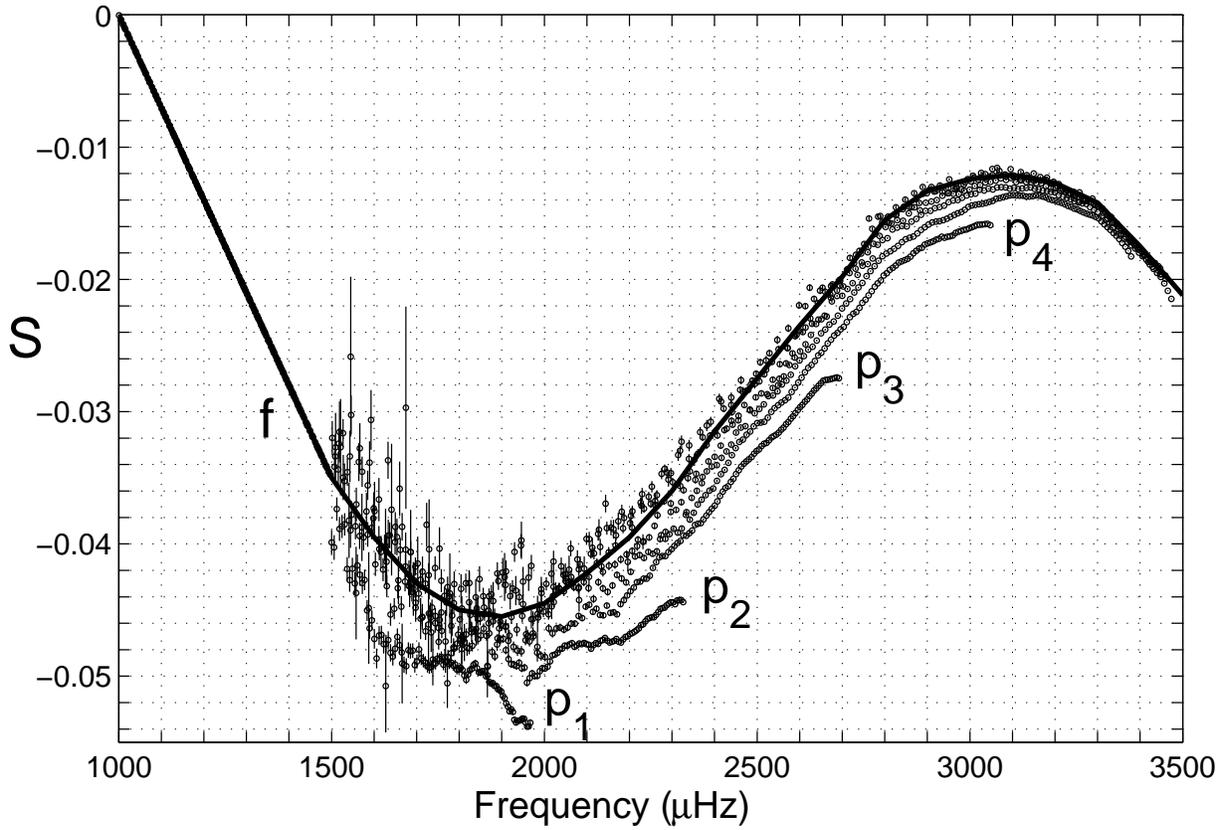} \caption{
 Same as Fig. 2, but for asymmetry parameter. The asymmetry parameter $S$ is extrapolated by a straight line at frequencies below 1500 $\mu$Hz, where its measurement suffers from uncertainties coming from smaller signal-to-noise ratio and narrow line profiles.
 \label{fig4}}
 \end{figure}

 \begin{figure}
 \epsscale{1}
 \plotone{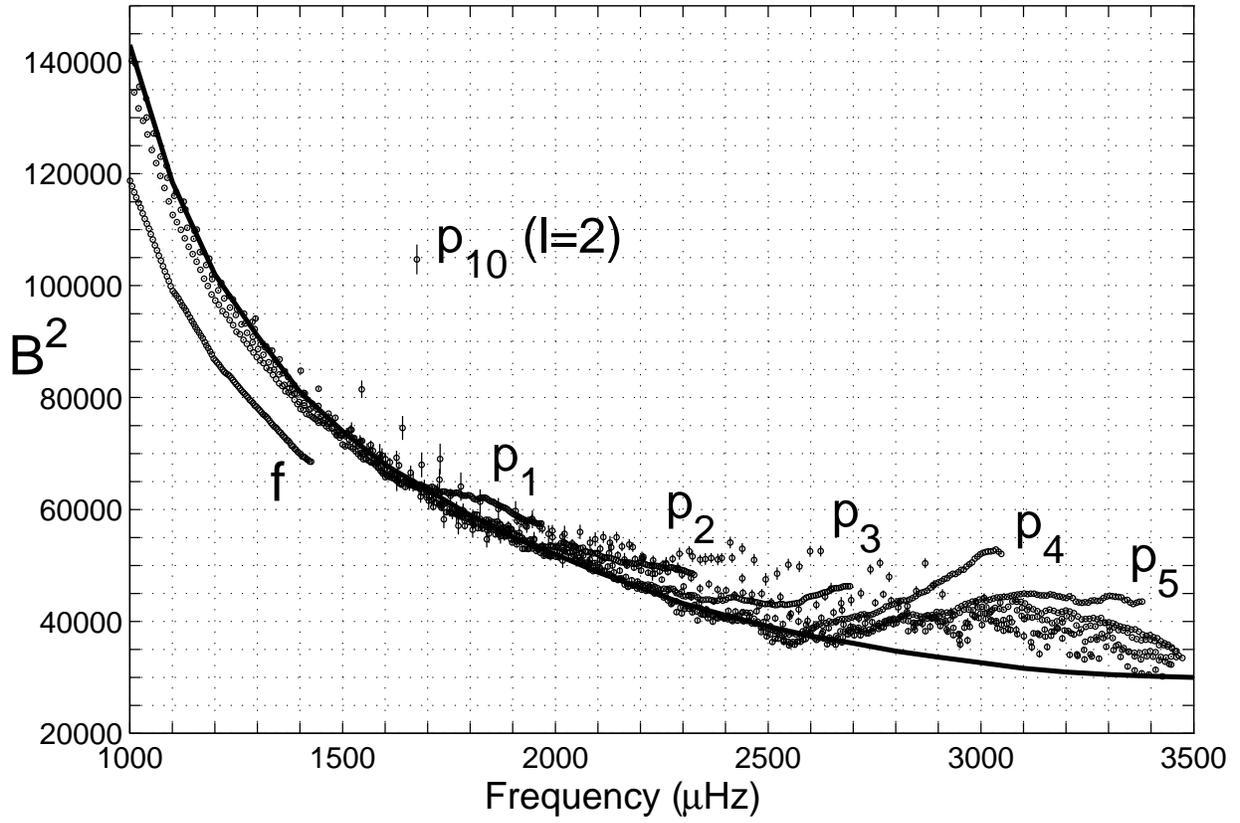} \caption{
 Same as fig. 2, but composite background.
 \label{fig5}}
 \end{figure}
 \begin{figure}
 \epsscale{1}
 \plotone{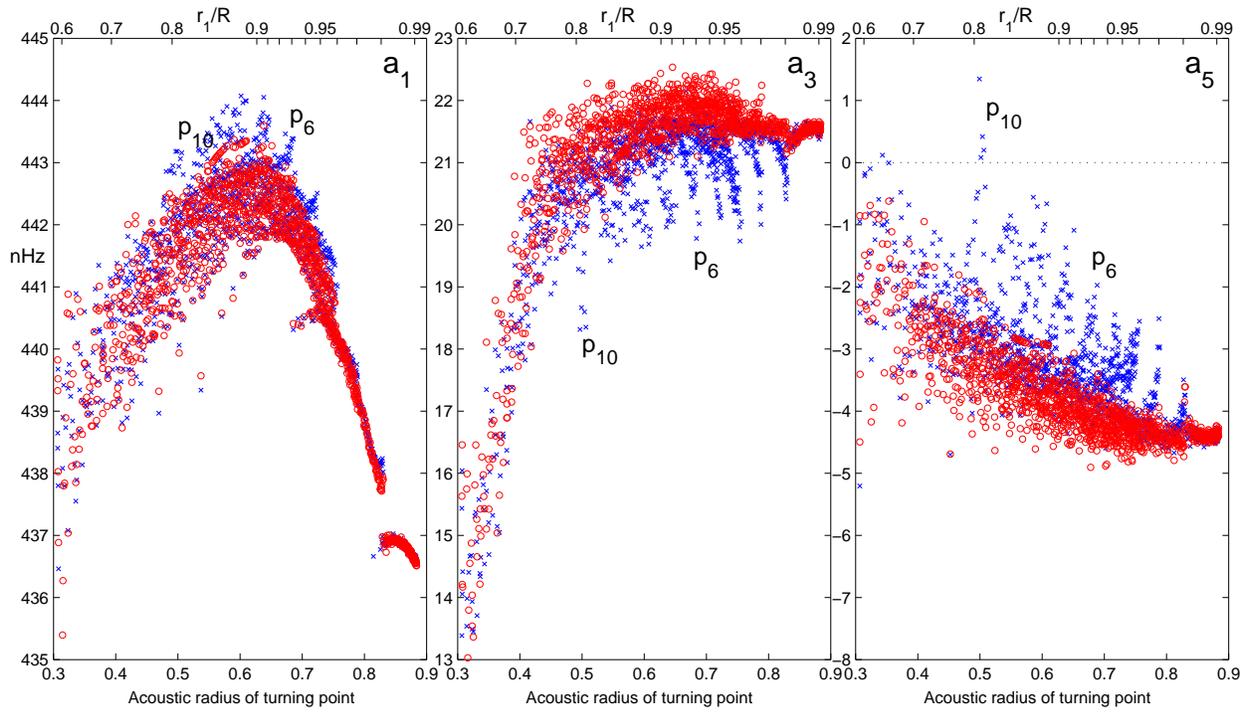} \caption{
 Rotational splitting coefficients $a_1, a_3, a_5$, inferred from the first year of SOHO MDI measurements (red circles). Blue crosses show the published coefficients {\bf \citep{schou99}}.
 \label{fig6}}
 \end{figure}
 \begin{figure}
 \epsscale{1}
 \plotone{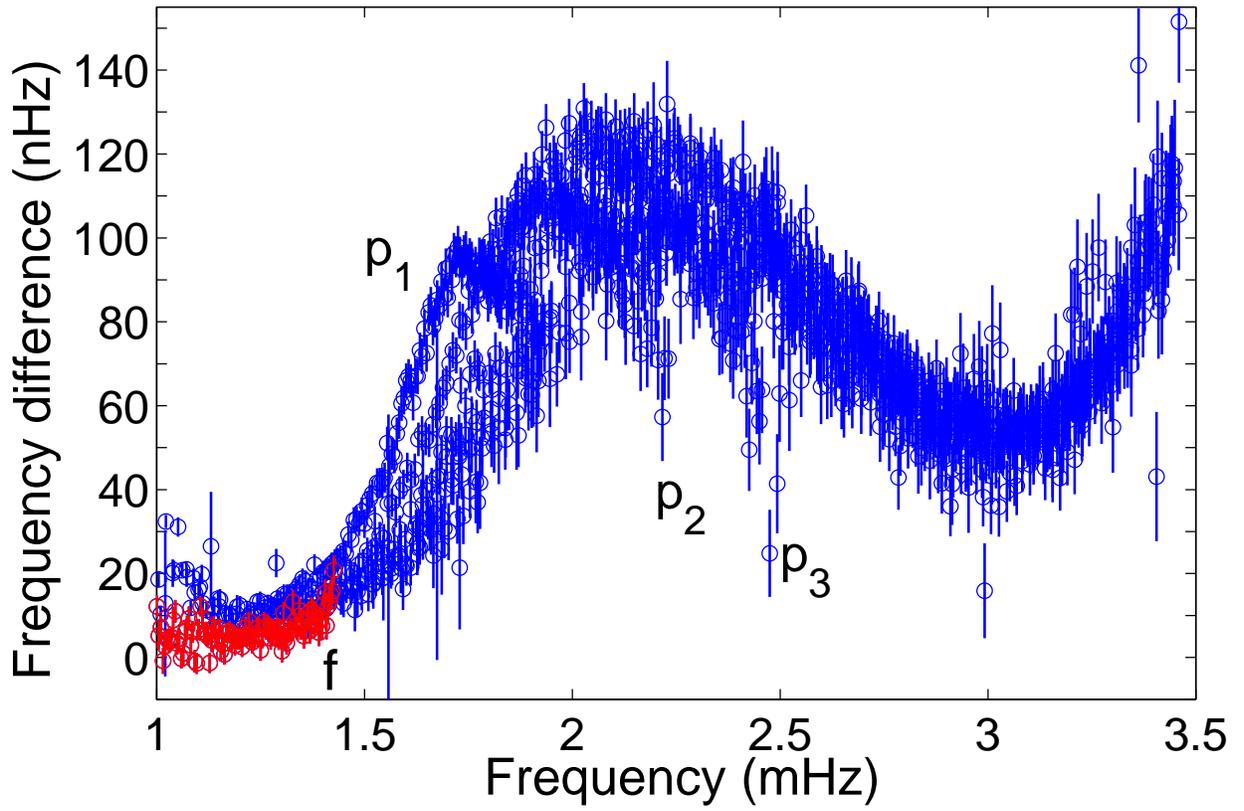} \caption{
 Difference between the centroid ($m$-averaged) frequencies and their published values {\bf \citep{schou99}}.
 \label{fig7}}
 \end{figure}

 \clearpage


\begin{thebibliography}{}
 \bibitem[Anderson et al.(1990)]{anderson90}
 Anderson, E. R., Duvall, T. L., Jr, \& Jefferies, S. M. 1990, \apj, 364, 699

 \bibitem[Basu et al.(2003)]{basu03} Basu, S., Christensen-Dalsgaard, J.,
 Howe, R., Schou, J., Thompson, M. J., Hill, F., \& Komm, R.
 2003, \apj, 591, 432

 \bibitem[Cox(1980)]{cox80}
 Cox, J. P. 1980, Theory of Stellar Pulsation (Princeton: Princeton Univ. Press)

 \bibitem[Duvall et al. (1993)]{duvall93}
 Duvall, T. L., Jr., Jefferies, S. M., Harvey, J. W., Osaki, Y.,
 \& Pomerantz, M. A. 1993, \apj, 410, 289

 \bibitem[Gough \& Vorontsov(1995)]{gv95}
 Gough, D. O., \& Vorontsov, S. V. 1995, \mnras, 273, 573

 \bibitem[Hindman(2011)]{hindman11}
 Hindman, B. W. 2011, arXiv:1112.4790v1 [astro-ph.SR] 20 Dec 2011

 \bibitem[Jefferies \& Vorontsov(2004)]{smj04}
 Jefferies, S. M., \& Vorontsov, S. V. 2004, \solphys, 220, 347

 \bibitem[Korzennik (2005)]{korz05} Korzennik, S. G.,
 2005, \apj, 626, 585

 \bibitem[Korzennik et al.(2004)]{korz04} Korzennik, S. G.,
 Rabello-Soares, M. C., \& Schou, J. 2004, \apj, 602, 481

 \bibitem[Korzennik et al.(2013)]{korz13} Korzennik, S. G.,
 Rabello-Soares, M. C., Schou, J., \& Larson, T. P. 2013, \apj, 772:87

 \bibitem[Larson \& Schou(2008)]{larson08}
 Larson, T. P., \& Schou, J. 2008, J. of Physics: Conf. Series, 118, 012083

 \bibitem[Nigam \& Kosovichev (1998)]{nigamkosovichev}
 Nigam, R., Kosovichev, A. G. 1998, \apj, 505, L51

 \bibitem[Nigam et al.(1998)]{nigam}
 Nigam, R., Kosovichev, A. G., Scherrer, P. H., \& Schou, J. 1998,
 \apj, 495, L115

 \bibitem[Rhodes et al.(2001)]{rhodes01}
 Rhodes, E. J., Jr., Reiter, J., Schou, J., Kosovichev, A. G., \& Scherrer, P. H. 2001,
 \apj, 561, 1127

\bibitem[Roxburgh \& Vorontsov(1995)]{rv95}
 Roxburgh, I. W., \& Vorontsov, S. V. 1995, \mnras, 722, 850

 \bibitem[Roxburgh \& Vorontsov(1997)]{rv97}
 Roxburgh, I. W., \& Vorontsov, S. V. 1997, \mnras, 292, L33

 \bibitem[Schou(1999)]{schou99}
 Schou, J. 1999, \apj, 523, L181

 \bibitem[Schou et al.(2002)]{schou02}
 Schou, J., Howe, R., Basu, S., Christensen-Dalsgaard, J., Corbard, T.,
 Hill, F., Komm, R., Larsen, R. M., Rabello-Soares, M. C., \& Thompson, M.J.
 2002, \apj, 567, 1234

 \bibitem[Schou (2013)]{schou13}
 Schou, J. 2013, Private Communication

 \bibitem[Varshalovich et al (1988)]{varshalovich}
 Varshalovich, D. A., Moskalev, A. N., \& Khersonskii, V. K., 1988,
 Quantum Theory of Angular Momentum, Singapore: World Scientific Publishing

 \bibitem[Vorontsov(2002)]{svv02a} Vorontsov, S. V. 2002,
 in:  From Solar Min to Max: Half a Solar Cycle with SOHO:
 Proc. {\it SOHO} 11 Symposium (ESA SP-508; Noordwijk: ESA), 107

 \bibitem[Vorontsov(2007)]{svv07} Vorontsov, S. V. 2007,
 MNRAS, 378, 1499

 \bibitem[Vorontsov \& Jefferies (2005)]{vj05}
 Vorontsov, S. V., \& Jefferies, S. M., 2005, \apj, 623, 1202 (Paper
 I)

 \bibitem[Vorontsov et al.(2002)]{svv02b} Vorontsov, S. V.,
 Christensen-Dalsgaard, J., Schou, J., Strakhov, V. N., \&
 Thompson, M. J.  2002,
 in:  From Solar Min to Max: Half a Solar Cycle with SOHO:
 Proc. {\it SOHO} 11 Symposium (ESA SP-508; Noordwijk: ESA), 111

 \bibitem[Vorontsov et al.(2009)]{svv09}
 Vorontsov, S. V., Jefferies, S. M., Giebink, C. and Schou, J. 2009,
 in: Solar-Stellar Dynamos as Revealed by Helio- and Asteroseismology,
 eds. M. Dikpati, T. Arentoft, I. Gonzalez Hernandez, C. Lindsey, and F. Hill,
    Astron. Soc. Pacific Conf.  Series, 416, 301

 \bibitem[Vorontsov et al.(2013)]{svv13}
 Vorontsov, S. V., Baturin, V. A., Ayukov, S. V., Gryaznov, V. K., 2013,
 MNRAS 430, 1636



\end{thebibliography}
\end{document}